\documentclass[12pt]{article}

\usepackage[margin=1in]{geometry}
\usepackage{CJKutf8, amsthm, graphicx, hyperref, orcidlink}

\theoremstyle{definition}
\newtheorem{assumption}{Assumption}

\usepackage[style=trad-unsrt, citestyle=numeric-comp, giveninits=true, doi=false, url=false, eprint=false, isbn=false]{biblatex}

\begin{document}

\title{Perfect score on IPhO 2025 theory by Gemini agent\footnote{Project repository: \url{https://github.com/yichenhuang/IPhO2025}}}

\author{Yichen Huang (黄溢辰)\orcidlink{0000-0002-8496-9251}\\
Santa Clarita, California 91350, USA\\
\href{mailto:huangtbcmh@gmail.com}{huangtbcmh@gmail.com}}

\begin{CJK}{UTF8}{gbsn}

\maketitle

\end{CJK}

\begin{abstract}

The International Physics Olympiad (IPhO) is the world's most prestigious and renowned physics competition for pre-university students. IPhO problems require complex reasoning based on deep understanding of physical principles in a standard general physics curriculum. On IPhO 2025 theory problems, while gold medal performance by AI models was reported previously, it falls behind the best human contestant. Here we build a simple agent with Gemini 3.1 Pro Preview. We run it five times and it achieved a perfect score every time. However, data contamination could occur because Gemini 3.1 Pro Preview was released after the competition.

\end{abstract}

\section{Introduction}

The International Physics Olympiad (IPhO) is the world’s most prestigious and renowned physics competition for pre-university students. Established in Poland in 1967 with just five participating countries, it has since been held annually and expanded to include over 100 nations, each represented by a team of up to five of their best pre-university students. The competition is a test of both broad physical skills and sustained concentration: over two consecutive days, contestants are given two 5-hour sessions to solve theoretical and experimental problems, respectively. The problems are drawn from a standard general physics curriculum. The curriculum is for freshman undergraduate students majoring in physics and covers the fields of mechanics, electromagnetism (including optics), thermodynamics, and special relativity.

Qualifying for the IPhO is extremely challenging. In the United States, a student must advance through a series of national competitions of increasing difficulty, from the $F=ma$ exam to the USA Physics Olympiad (USAPhO). Top performers at USAPhO are invited to compete for the five spots on the U.S. national team. Most other countries have similarly stringent selection processes in order to find their best pre-university students to represent the country. A gold medal at the IPhO is an extraordinary achievement, awarded to only the top one-twelfth of the contestants. Thus, it deserves to be permanently highlighted on the curriculum vitae.

Advanced reasoning capability is a core form of intelligence. Consequently, automated reasoning has become a major frontier of artificial intelligence (AI). A lot of effort is devoted to improving the reasoning capability of AI models, especially large language models (LLM). The progress is so fast that we have to evaluate the models on benchmarks of increasing difficulty. ``Easier'' mathematical and scientific datasets like the American Invitational Mathematics Examination (AIME), GPQA \cite{RHS+24} have already been almost saturated by the state-of-the-art models. The next tier is Olympiad-level problems, which require correct final answers supported by sound scientific arguments. For almost every discipline, gold-medal but unsaturated performance on the corresponding international Olympiad in 2025 has been reported; see Table \ref{tab:1}. AI models can even sometimes solve research-level problems \cite{ZTY+25, WCJ+26, ABH+26}.

\begin{table}
\caption{Gold-medal but unsaturated performance on the international Olympiad of various disciplines in 2025. For IMO 2025, we only cite the first batch of announcements made soon after the event. For IPhO and IChO, only theoretical problems are evaluated. The numbers in smaller fonts are obtained by evaluating solutions with LLM and may not be accurate.}
\label{tab:1}
\begin{tabular}{c|c}
\hline
event & accuracies \\ 
\hline
International Mathematical Olympiad (IMO) & 83.3\% \cite{WSB25, LL+25, HY25, BDS25}\\
\hline
International Physics Olympiad (IPhO) & 78.3\% \cite{QSJ+25}, {\scriptsize75.7\%} \cite{YWC+25}, {\scriptsize71.7\%} \cite{YYW+25}, {\scriptsize77.3\%} \cite{CCY+25},\\
& {\scriptsize72.7\%, \scriptsize84\%} \cite{LWC+26}, {\scriptsize87.7\%} \cite{DT26}, 100\% [this work] \\
\hline
International Chemistry Olympiad (IChO) & {\scriptsize75.4\%} \cite{XBC+25}, {\scriptsize82.8\%} \cite{DT26}\\
\hline
International Olympiad in Informatics (IOI) & 74.4\% \cite{SFN+25}, 88.9\% \cite{H+25} \\
\hline
International Olympiad on & 85.6\% (theory), 88.5\% (data analysis) \\
Astronomy and Astrophysics (IOAA) & \cite{PCP+25} \\
\hline
\end{tabular}
\end{table}

While there are more people working on AI for mathematics, physics is also an important and the most fundamental discipline of natural science. Mastering IPhO problems is one step towards making AI more useful in physics. In this paper, we build a simple agent powered by Gemini 3.1 Pro Preview. The agent synthesizes solutions from parallel thinking \cite{DLT24} (analogous to Gemini 2.5 Deep Think \cite{GT25} and Grok 4 Heavy) and contains a component that may use Python code to make measurements on figures \cite{SMV23}. We run the agent $5$ times on IPhO 2025 theory problems. It achieved full scores every time. To our knowledge, this is the first time perfect performance is observed in an international Olympiad event.

Gemini 3.1 Pro Preview was released after IPhO 2025. Thus, IPhO 2025 problems could occur in the training dataset of the model, and the perfect performance by our Gemini agent should be interpreted with caution. However, it should be clear that our result is still meaningful. The second highest reported result 87.7\% was achieved by Gemini 3 Deep Think \cite{DT26}. In the release note \cite{GT26}, the Gemini team indicates that Gemini 3.1 Pro is the ``upgraded core intelligence that makes [the Gemini 3 Deep Think] breakthroughs possible.'' Thus, the risk of data contamination for Gemini 3 Deep Think is at the same level as that for our agent built on Gemini 3.1 Pro Preview.

The repository (\url{https://github.com/yichenhuang/IPhO2025}) of this paper contains
\begin{itemize}
\item problem statements (PDF documents) downloaded from the official website and their pre-processed and corrected version to be fed into the agent;
\item official solutions and marking schemes;
\item code of our agent, with all prompts;
\item five solutions of each problem generated by the agent.
\end{itemize}

\section{Related work}

The IPhO 2025 theoretical exam has three problems. Each is worth 10 points so that the full score is 30. Each problem is divided into 3 to 4 parts, and each part has 3 to 5 sub-problems.

There are already many evaluations of LLM on IPhO 2025 in the literature. All results below are reported by the raw points (out of 30) rather than the accuracies in percentages.

Qiu et al.~\cite{QSJ+25}, where the second named author is an IPhO gold medalist, used Gemini 2.5 Pro only. The model alone achieved 21.4 points. They also built the agent ``Physics Supernova'' using smolagents, which is an implementation of ReAct \cite{YZY+23}. The agent achieved 23.5 points. It can be applied broadly to other problems because the ReAct framework is general (not tailored to physics). This reference stands out because it uses human evaluation. (All other results on IPhO 2025 discussed in this section are obtained by automatic evaluation.)

References \cite{YWC+25, YYW+25, CCY+25, LWC+26} have largely overlapping sets of authors. HiPhO \cite{YWC+25} is mostly a dataset paper but evaluated a broad set of models on various physics Olympiads. On IPhO 2025, the best result is 22.7 points by Gemini 2.5 Pro. Reference \cite{YYW+25} built the agent ``PhysicsMinions.'' Powered by Gemini 2.5 Flash, it achieved 21.5 points. P1 \cite{CCY+25} is an open weight model trained from Qwen3. P1-235B-A22B achieved 21.2 points alone and 23.2 points with PhysicsMinions. Note that P1 is a purely textual model. For IPhO problems that contain figures, it can only make an educated guess of the problem setup or give up. Finally, P1-VL \cite{LWC+26} is an open weight model trained from Qwen3-VL. P1-VL-235B-A22B achieved 21.0 points alone and 22.1 points with PhysicsMinions. While P1-VL has visual capabilities, its performance on IPhO 2025 is worse than that of P1. We suspect that this is because the figures in IPhO 2025 problems are too challenging to understand visually. Misunderstanding figures might hurt the model's performance.

This line of work \cite{YWC+25, YYW+25, CCY+25, LWC+26} uses Gemini 2.5 Flash to automatically grade the solutions generated by various models. The evaluator is given the official marking scheme. The final score for each sub-problem is the greater of the answer-level score and the step-level score. Thus, a solution with a correct final answer always receives full credit while that with an incorrect answer is graded by the number of correct intermediate steps. This grading convention leads to inflated scores if the phenomenon of correct answers with incorrect reasoning occurs. This phenomenon occurs frequently in LLM for mathematics \cite{ZZZ+25} and could also occur when applying LLM to physics. While being good at grading final answers, even the state-of-the-art LLM cannot accurately grade the intermediate reasoning in a solution to a technically challenging problem, even if the official marking scheme is given \cite{LHN+25}. Indeed, Table 7 in Ref.~\cite{YWC+25} presents a comparative study, which shows that Gemini 2.5 Flash cannot accurately reflect human grading. In our opinion, automatic evaluation by LLM definitely correlates positively with human grading, but human grading is needed for accurate results.

In a very recent evaluation, Gemini 3 Deep Think achieved 26.3 points, where answer correctness was graded with reference to canonical solutions using Gemini as a judge \cite{DT26}.

Reference \cite{JLY+25} claimed nearly perfect performance on Chinese Physics Olympiad theory problems by an agent built from Gemini 2.5 Pro.

\section{Dataset}

\subsection{Data collection and pre-processing}

We downloaded the problem statements (PDF documents) from the official website (\url{https://www.ipho2025.fr/official-questions-ipho-france-2025}). Since there is no guarantee that the website will be permanently accessible or that the PDFs for problem statements will not be silently updated, we include a copy of the three PDFs in our project repository for archival purposes.

Unlike IMO problems which are usually one or two short paragraphs of text, IPhO problems are long and multimodal: The three IPhO 2025 problems have an average of about 5 pages and 5 figures. There is some freedom in how to input the problem statements to a model or an agent, and the final accuracy depends on the input format. Thus, we believe that it is important to be fully transparent on how we pre-processed the dataset because pre-processing may significantly affect the final score. The most convenient way is to directly feed in the PDF files (i.e., no pre-processing), and the model would use OCR to automatically extract the text and figures. This approach is bad because OCR frequently makes errors, which downgrade the final performance. We should manually convert the PDFs in order to reduce the possibility that the problem statement is misunderstood. However, we must not edit the semantics or even accidentally provide any hint during pre-processing.

We manually extracted the text and, following Ref.~\cite{QSJ+25}, converted it into markdown format (i.e., saved into an md file). We also added the mark ``---'' to separate the sub-problems. These marks make it easy for the code to parse the sub-problems when reading the md file. They are not fed into the model.

We extracted the figures by first converting the PDF into a high-quality png and then cropping the png. Some figures in the PDF are vector graphs, whose resolution can in principle be made arbitrarily high, but the resolution of raster graphs is capped.

Certain figures in the PDF have large background area. For humans, it is very clear that background is meaningless. However, Gemini spends a fixed number of tokens representing a figure. Background area consumes tokens and thus there are fewer ``useful'' tokens describing the key elements in the figure. To reduce this token loss because of the background, we crop the figure by taking the minimum region (with reasonable margins) that contains the object or system of interest. Indeed, focusing on the area of interest by cropping out irrelevant regions is a well-known pre-processing technique to improve image understanding.\footnote{I thank Pengchuan Zhang for pointing this out.}

Problem 3 has 6 figures. Each of the first three has one panel and each of the last three has two panels. The two panels in the same figure are not closely related to each other. They form a single figure so that they can be placed side by side on a page in order to reduce the white space to their left and right. We split the two-panel figures into, e.g., Fig.~4A and Fig.~4B. After the splitting, each figure has a unique ID, which is either a pure number or a number followed by A or B. The existence of a letter in the ID means that the figure was a panel in a two-panel figure. We also split all two-panel figures in Problems 1 and 2. Since Gemini spends a fixed number of tokens per figure, splitting a two-panel figure into two doubles the total number of tokens for the same visual content. This may potentially improve the accuracy of image understanding.

While we described our figure pre-processing as cropping and splitting, in practice, all we did is taking snapshots of the PDF for problem statements. Cropping means taking a snapshot with a smaller rectangular bounding box, and splitting means taking two snapshots (one for each panel) on a two-panel figure. When figures in a dataset are obtained by taking snapshots, the dataset preparer has to define a rule for constructing bounding boxes. We believe that any reasonable rule is acceptable as long as it is consistently applied and does not edit the content of any figure.

After extracting the figures in the PDF into png files, we inserted them into the textual md file. The official Gemini documentation by Google recommends that figures be placed before the textual descriptions of them. Thus, we put the figure files immediately before the paragraph (in the md file) that first mentions the figure.

Each IPhO problem also has an answer sheet as a standalone PDF file, which may contain figures and/or tables. Sometimes the problem statement refers to the answer sheet. For example, Sub-problem A.2 in Problem 2 reads ``..., complete the table in the answer sheet to ...'' In this case, we have to incorporate the table into the problem statement because otherwise the problem statement is not self-contained so that the model would not know what to do. We also manually added a small piece of text ``The table on the answer sheet to be completed:'' to clarify that the table is from the answer sheet. If the problem statement refers to the answer sheet but is already complete without the answer sheet, we do not incorporate tables or figures from the answer sheet, e.g., Sub-Problem B.1 in Problem 2. Thus, we made minimal edits to the problem statement because of the answer sheet.

The above is the full description of how we pre-processed the dataset. We have not done ablation study on to what extent or even whether each of the pre-processing operations affects the final accuracy. With all pre-processing, our agent can achieve perfect scores. Missing a single pre-processing operation would not crush the performance of the agent. It could at most slightly reduce the accuracy, and the effect (if any) can only be observed by sampling a large number of solutions using the agent. This is not only expensive but also requires a lot of human labor because we have to use human grading for accurate evaluation.

As of February 2026, the official solutions or marking schemes are still not available on the official website, but they appear at \url{https://ipho.olimpicos.net}. References \cite{QSJ+25, YWC+25} used the official solutions and marking schemes on this unofficial website. We also use them.

\subsection{Curation}

We must make sure that the dataset is completely error-free before using it. To this end, we used Gemini 3.1 Pro Preview to review the problem statements and official solutions. It found a few errors, which are really errors upon our human evaluation. This subsection explains each error and how we have edited the problem statement or marking scheme accordingly. All our experiments were run and evaluated on the corrected dataset. Surprisingly, we do not find any of these errors being pointed out publicly on the Internet or in previous works listed in the IPhO row of Table \ref{tab:1}. Catching these errors is the physics contribution of this work. The resulting corrected dataset with the pre-processing described in the preceding subsection is available in our project repository. It facilitates future testing of other models on corrected IPhO 2025 problems.

As LLMs become stronger in reasoning, they are evaluated on increasingly more advanced problems. The errors we found show the importance of having a domain expert overseeing the evaluation.

\subsubsection{Fig.~1(B) in Problem 1}

\begin{figure}
    \centering
    \includegraphics[width=.5\textwidth]{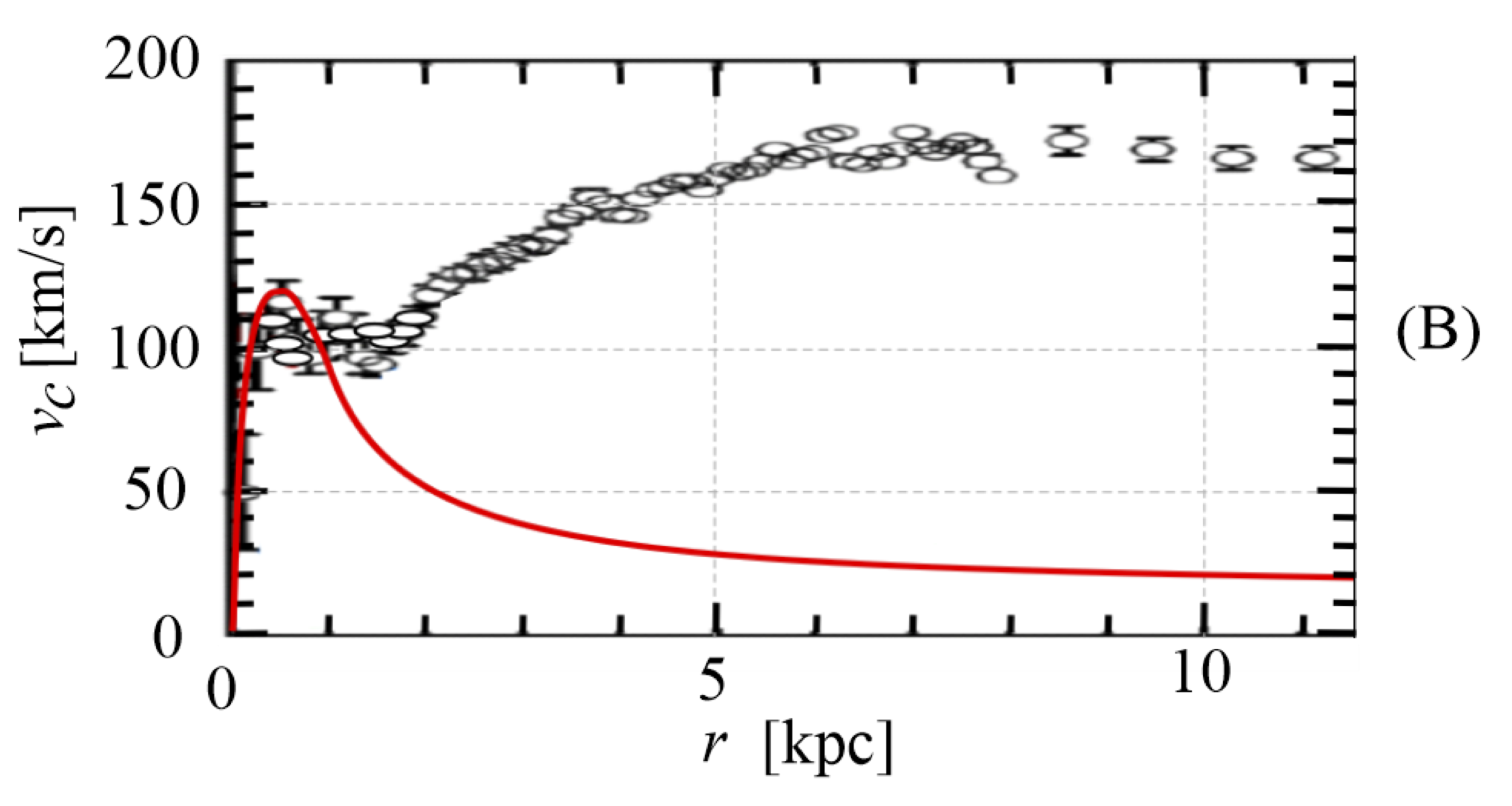}
    \includegraphics[width=.49\textwidth]{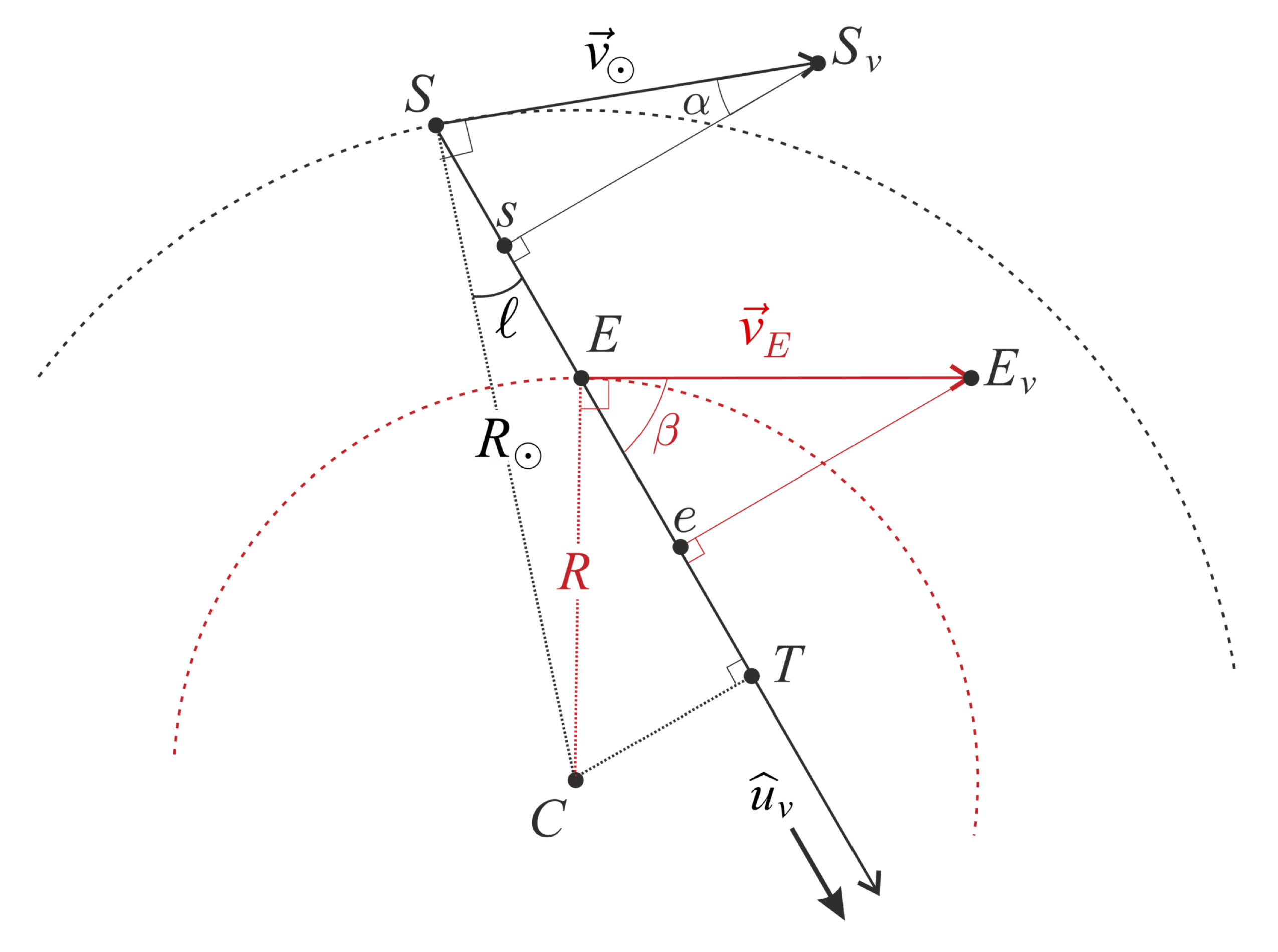}
    \caption{Left panel: Fig.~1(B) in Problem 1; Right panel: Fig.~2 in Problem 1.}
    \label{fig}
\end{figure}

Part B of Problem 1 considers a spherical galaxy whose mass distribution is spherically symmetric around the galactic center. Sub-problem B.2 further assumes the following:
\begin{assumption} \label{ass:1}
The mass distribution is roughly homogeneous in a central region named the bulge, and there is almost no mass outside the bulge.
\end{assumption}
The red curve in Fig.~1(B) in Problem 1 (see Fig.~\ref{fig}, left panel) predicts the speed $v_c$ of an object with a perfect circular orbit as a function of the distance $r$ to the galactic center. The sub-problem asks to deduce the mass $M_b$ of the bulge by reading values on the red curve.

The official solution correctly finds that $v_c(r)=\sqrt{GM_b/r}$ for $r>r_b$, where $G$ is the gravitational constant, and $r_b$ is the radius of the bulge. It is not explicitly stated but follows from Assumption \ref{ass:1} that $r_b$ corresponds to the peak of the red curve. However, the red curve on the right of the peak cannot be fitted by $C/\sqrt r$ for some constant $C$; it decays faster than $1/\sqrt r$ as $r$ increases. This is a fundamental physics error (not just an inaccuracy) because in any spherical galaxy (possibly violating Assumption \ref{ass:1}), the fact that mass cannot be negative implies $v_c(r_1)\sqrt{r_1}\le v_c(r_2)\sqrt{r_2}$ for any $0<r_1<r_2$. A proof of this inequality can be obtained by querying Gemini 3.1 Pro Preview.

A contestant is supposed to read a data point from the red curve and numerically calculate
\begin{equation} \label{eq:Mb}
M_b=v_c^2r/G.
\end{equation}
Line B.2.5 in the official marking scheme estimates that $M_b\simeq9\times10^8M_\odot$ and explicitly allows a tolerance of $\pm25\%$ for the numerical value of $M_b$. The contestant can choose any $r$ on the right of the peak. Because of the aforementioned error, different $r$ leads to different results. If the chosen $r$ is small, the calculated $M_b$ is larger than $125\%\times9\times10^8M_\odot$, even if the reading of the data point is accurate.

For this error, we do not modify the problem statement or Fig.~1(B), but modify the marking scheme as follows. For Line B.2.4, the contestant can choose any data point on the right of the peak. For Line B.2.5, full points (0.1) are awarded as long as the relative error with respect to the value of $M_b$ calculated using the precise reading of the chosen data point is $\le25\%$. This revised marking scheme still requires the contestant to read the data point relatively accurately and apply the correct formula (\ref{eq:Mb}).

\subsubsection{Fig.~3 in Problem 1} \label{ss:err}

Figure 2 in Problem 1 (see Fig.~\ref{fig}, right panel) gives the correct geometry of the measurement. The derivation in the official solution to Sub-problem C.3 yields the correct relationship
\begin{equation}
v_{rE/S}=v_\odot(R_\odot/R-1)\sin\ell.
\end{equation}
Here, $v_{rE/S}$ is the relative velocity of the emitter with respect to the sun in the direction of sight; $v_\odot$ is the speed of the sun in its orbit around the center $C$ of the Milky Way; $R_\odot$ is the distance from the sun to $C$; $R$ is the distance from the emitter to $C$; $\ell=30^\circ$ is the galactic longitude.

Figure 2 in Problem 1 clearly shows $R<R_\odot$ so that $v_{rE/S}>0$. Positive $v_{rE/S}$ means that the emitter is moving away from the sun. This leads to redshift (frequency becomes smaller). However, all three peaks in Fig.~3 in Problem 1 have blueshift $f-f_0>0$, where $f$ and $f_0$ are the observed and original frequencies, respectively. Thus, Figs.~2 and 3 in Problem 1 contradict each other.\footnote{I thank Lin F. Yang for independently confirming this error.}

To fix this error, we manually change the horizontal axis in Fig.~3 of Problem 1 from $f-f_0$ to $f_0-f$. Line C4.1 in the marking scheme is changed to $v_{r,i}=-c\Delta f_i/f_0$. No further changes to the problem statement or official solution are necessary. Note that the official solution to Sub-problem C.5 also depicts $R<R_\odot$, yielding redshift.

\subsubsection{Sub-problem B.3 in Problem 3}

The statement of this sub-problem does not contain any error, but there is a miscalculation in the official solution. Line B.3.5 reads
\begin{equation}
h=\frac{2\sigma}{\rho_\ell v_f^2}=\frac{3t_b^2}{2a^4}\sqrt{\frac{\sigma^3}{\rho_\ell^3g_0}}.
\end{equation}
The first equality is correct (quoted from Line B.1.6) but the second is not. Line B.3.4 gives $v_f=\frac{r_c}{t_b}$ and $r_c=\frac2{\sqrt3}a^2\sqrt{\frac{\rho_\ell g_0}\sigma}$. Correct calculation yields
\begin{equation} \label{eq:b3}
h=\frac{2\sigma}{\rho_\ell v_f^2}=\frac{2\sigma t_b^2}{\rho_\ell r_c^2}=\frac{3\sigma^2t_b^2}{2a^4\rho_\ell^2g_0}.
\end{equation}
The numerical value of $h$ in Line B.3.6 is calculated using the correct formula and is correct.

For this error, we do not modify the problem statement or the marking scheme, but use the correct formula (\ref{eq:b3}) in human evaluation.

\section{Agent}

Problem 2 in IPhO 2025 does not require making measurements on figures but Problems 1 and 3 do. We describe our agent for Problem 2 in Subsection \ref{ss:2}. Then, we describe the component for making measurements and how to incorporate it into the agent in Subsection \ref{ss:13}. The code for our agent is available in our project repository.

Each IPhO problem has many sub-problems, which largely form a logical progression. A sub-problem may depend on the solutions to some of the previous sub-problems. To use an LLM (not an agent) to solve a problem, the most natural scheme is to have a multi-round conversation. The system instruction describes the task as solving an IPhO problem. Then, the user inputs the sub-problem statements one by one. When solving each sub-problem, the model can refer to all preceding context including its own solutions to previous sub-problems. More specifically, we index the sub-problems by natural numbers starting from $1$. At turn $n$ of the conversation, the user's message is the statement $q_n$ of Sub-problem $n$, and the model's response $a_n$ is the solution of the sub-problem. Thus, we have the sequence
\begin{equation} \label{eq:seq}
q_1,a_1,q_2,a_2,q_3,\ldots,q_{n-1},a_{n-1},q_n,a_n,\ldots
\end{equation}

This scheme makes the maximum use of the context and best mimics how a human contestant would tackle the problem. However, if the model makes even a minor error in an early sub-problem, the error may be severely penalized because it provides misleading context for all subsequent sub-problems. Thus, the sample-to-sample fluctuation of the accuracies may be large. One has to run the model a sufficient number of times for a precise estimation of the expected accuracy and the variance.

\subsection{Agent for Problem 2} \label{ss:2}

Our agent for Problem 2 works as follows. Let $H_m$ be the dialogue history consisting of the first $m$ messages in the sequence (\ref{eq:seq}) such that the sequence can be written as $H_{2n-2},q_n,a_n,\ldots$ or $H_{2n-1},a_n,\ldots$. We independently generate four raw solutions $a_n^{(1)},a_n^{(2)},a_n^{(3)},a_n^{(4)}$ to Sub-problem $n$ from the context $H_{2n-1}$. Then, we synthesize $a_n^{(1)}$ and $a_n^{(2)}$ such that $a_n^{(5)}$ is the output of $H_{2n-2},q_n+a_n^{(1)}+a_n^{(2)}+$[synthesis prompt], where the plus sign denotes concatenation. The synthesis prompt asks the model to review the two provided solutions, identify errors (if any) in them, and construct a correct solution. Similarly, we synthesize $a_n^{(3)}$ and $a_n^{(4)}$ into $a_n^{(6)}$, and synthesize $a_n^{(5)}$ and $a_n^{(6)}$ into $a_n$. Then, we move on to the next sub-problem with the new context $H_{2n}=(H_{2n-1},a_n)$.

It should be clear that we are just using the standard idea of synthesizing solutions from parallel thinking \cite{DLT24} (analogous to Gemini 2.5 Deep Think \cite{GT25} and Grok 4 Heavy). While this idea is usually advertised as exploring diverse solution paths, here its main functionality is error correction. LLM can make mistakes. Then, identifying the incorrect solution between two conflicting solutions is easier than determining whether a given solution is incorrect.

Our agent for Problem 2 generates four raw solutions per sub-problem and synthesizes them in two rounds. A lighter agent that only synthesizes two raw solutions per sub-problem would perform worse but still reasonably well, at less than half the cost. We estimate that the probability of the light agent achieving a perfect score on Problem 2 is about 90\%, but have not gathered sufficient data to justify this estimate.

In either agent for Problem 2, we disable the code execution tool because Gemini 3.1 Pro Preview can perform numerical calculations reliably.

\subsection{Agent for Problems 1 and 3} \label{ss:13}

Some sub-problems in Problems 1 and 3 require making accurate measurements on figures. However, Gemini 3.1 Pro Preview's native vision is not good at this task and can only make accurate measurements sometimes. Indeed, measuring with LLM is analogous to a human making measurements by eyeballing a figure. A human needs a ruler or other tools to achieve high measurement accuracy. Similarly, given the code execution tool, LLM can accurately measure distances on figures in pixels by writing a Python script \cite{SMV23} using standard computer vision (for object detection) and data processing libraries. For Gemini 3 Flash and later models, this capability is known as agentic vision \cite{RD26}.

The component for making measurements can be constructed and incorporated into our agent as follows. Suppose we have already solved the first $n-1$ sub-problems. Disabling Python, we first generate a solution $a_n'$ to Sub-problem $n$ from the context $H_{2n-1}$. Then, we send in $H_{2n-1},a_n'$,[measurement detection prompt]. The detection prompt asks the model to detect whether any measurement is made in $a_n'$, and restricts the output of the model to be either (1) ``yes,'' followed by a statement describing what quantities are measured on which figure in the solution $a_n'$ or (2) ``no.'' In the latter case, let $a_n=a_n'$, and we directly move on to the next sub-problem with the new context $H_{2n}=(H_{2n-1},a_n)$.

In the ``yes'' case, we call the measurement component, which is a standalone single-round conversation with Gemini 3.1 Pro Preview. The user's message (input) is the figure and the descriptions of the quantities to measure. The model's response (output) is the measurement results. While agentic vision (with Python) usually gives more accurate results than native vision (without Python), there are special cases where native vision is better. For example, when a data point exactly locates at a grid point, native vision can read the intended value without error while agentic vision may introduce an artificial error of, say, $1\%$. Thus, the model has to decide whether or not to use agentic vision and makes the measurement accordingly. Only within the measurement component the model may use Python.

We observe that the measurement component can make mistakes with small probability. For example, the object detector may identify an incorrect object so that the result is not even close to the ground truth. To make measurements more robustly, we measure the same quantity three times and take the median (not average).

Let $a_n$ be the output of $H_{2n-1},a_n'$,[update prompt]+[new measurement results]. The update prompt simply asks the model to rewrite the solution $a_n'$ using the new measurement results obtained from the measurement component. We move on to the next sub-problem with the new context $H_{2n}=(H_{2n-1},a_n)$.

\subsection{Discussions}

Both synthesizing solutions from parallel thinking and making measurements with Python code are standard ideas. Our contribution is to find the combination of methods that leads to perfect performance on IPhO 2025 theory problems. This result is beyond all previously reported ones. Since one does not expect that a universal agent leads to best results in all domains, designing domain-specific agents is a nontrivial task that pushes the frontiers of the capability of AI in a domain. The task also requires a deep understanding of the strengths and weaknesses of AI models in the domain.

An agent is a combination of a base model and an agentic workflow. If the base model is not strong enough (e.g., misunderstands a physical principle required to solve a problem), then the agent cannot achieve a perfect score regardless of how the workflow is designed. Otherwise, the stronger the base model is, the less test-time compute and simpler the workflow can be. Our agent only uses parallel thinking (and agentic vision) because Gemini 3.1 Pro Preview is a very strong model. Other base models may also be able to achieve a perfect score with a more sophisticated workflow that possibly contains the following techniques.

Consensus \cite{WWS+23} and verification-and-refinement are two standard inference-time methods. The former is very effective in boosting the probability of solving a problem towards $1$ if the problem can be solved with probability greater than $1/2$. The latter can substantially improve the capability of solving IMO problems \cite{HY25} and has already been implemented in earlier works \cite{QSJ+25, YYW+25} on building agents for IPhO.

\section{Evaluation}

\subsection{Settings}

Throughout this work, we use Gemini 3.1 Pro Preview with the default thinking level ``high'' and the default temperature $1$ (the latter is strongly recommended by official Gemini documents for reasoning tasks). We use ultra high media resolution. Each image in ultra high resolution consumes about 2240 tokens. Since some words in physics (e.g., ``force'') could be misinterpreted as violence, we turn off all safety settings. We use Files API for inputting figures to the model and disable web search.

Gemini 3 models have the feature ``thought signature,'' which is an encrypted representation of the model's internal thought process (chain of thought). Official Gemini documents strongly recommend turning on thought signatures in a multi-round conversation so that when generating later responses the model can attend to not only earlier responses but also its previous internal thinking. However, in an agentic workflow individual turns in a conversation can be deleted, replaced, or modified. Thought signatures might break the coherence of the conversation because they could refer to a deleted or replaced turn. Hence, we turn thought signatures off in our agent.

For Problem 2, we use the agent that generates four raw solutions per sub-problem and without the measurement component. For Problems 1 and 3, we use the agent that generates two raw solutions per sub-problem and with the measurement component. We could use the same agent (four raw solutions and with the measurement component) for all three problems. We make different choices for different problems only in order to save inference time and cost. This is not overfitting. Problems 1 and 3 are relatively easy so that two raw solutions suffice, but four raw solutions do not worsen the performance.

\subsection{Grading}

Throughout this work, we use human evaluation, which takes quite a bit of labor.

Some sub-problems (C.5 in Problem 1 and B.1, C.3 in Problem 2) require drawing graphs or diagrams. Since the output of Gemini 3.1 Pro Preview is text only, we added a system instruction requiring a mathematically precise textual description of how to draw the figure. The points for drawing are awarded if a student can unambiguously reconstruct the intended figure from the textual description.

IPhO marking schemes usually divide a solution into many small steps and assign partial points to each of them. This is nice to contestants because incremental progress is credited. However, sometimes essentially correct solutions skip one or more minor steps. We added a system instruction requiring maximally detailed presentation. This instruction significantly reduces the frequency of but cannot completely remove the phenomenon of step skipping. In our human evaluation, a solution is not penalized if the step it skips is minor to the extent that it is easy to fill by an expert. The implementation of this criterion is a bit subjective, but we expect that most physics professors would not penalize any of the five solutions generated by our agent for each problem.

For some sub-problems, the agent's method is very different from the official solution. In this case, full points are awarded as long as an expert can verify the correctness of the agent's solution without needing to fill in any gaps.

The agent is only required to solve tasks explicitly stated in a sub-problem. For example, Sub-problem D.4 in Problem 1 reads
\begin{quote}
Considering relevant cases, determine $v_c(r)$ for all values of $r$ in the MOND theory in the case of a gravitational field due to a homogeneously distributed mass $M$ with radius $R_b$.
\end{quote}
This sub-problem is worth 0.9 points. It should be clear that the sub-problem statement only requires deriving $v_c$ as a function of $r$. However, the marking scheme assigns 0.1 points for calculating the asymptotic behavior of $v_c(r)$ as $r\to\infty$ and another 0.1 points for the behavior of $v_c(r)$ as $r\to0$. All five solutions of our agent correctly derive $v_c(r)$ for all values of $r$, but do not always calculate the asymptotic behaviors. We do not penalize the solutions because calculating the asymptotic behaviors of $v_c(r)$ is not required by the sub-problem statement.

\subsection{Results}

We run our agent five times on each problem. All solutions of our agent are available in our project repository. Upon human evaluation, we found that every solution achieves a perfect score. While there are no errors in the final solutions, there are errors in the raw solutions from which the final solution is synthesized. In this subsection, we discuss these ``hidden'' errors that have all been corrected by our agentic workflow, but they prevent Gemini 3.1 Pro Preview from directly achieving a perfect score robustly.

We occasionally observed that Gemini 3.1 Pro Preview lost a constant pre-factor when copying a result from its own solution of a previous sub-problem, yielding a wrong final result for the current sub-problem. However, when synthesizing with a correct solution it is easy for the model to find the error. This type of error is so rare that we only observed once. While LLMs have been more and more reliable in algebraic manipulation, it is understandable that they occasionally make trivial mistakes because they are inherently random.

There are two related errors that appear much more often. We run our agent 5 times on Problem 2. In each run, 4 raw solutions were generated for each sub-problem. Thus, we have a total of 20 raw solutions for each sub-problem. All of them are generated conditioned on correct solutions to all previous sub-problems (because all final solutions achieve perfect scores). For both Sub-problems C.2 and C.3, 5 out of the 20 raw solutions are wrong.

The correct final result of Sub-problem C.2 is
\begin{equation} \label{eq:C2}
\vec T\simeq\frac{4S_bS_c}{S_b+S_c}\rho gX\vec{u_x}.
\end{equation}
Here, $\vec T$ is the tension force on a mass; $X$ is the displacement of the mass in the direction of the unit vector $\vec{u_x}$; all other symbols are constants. $\vec T$ is a destabilizing force and drives the mass away from the origin. In all 5 wrong raw solutions to C.2, the result has an extra minus sign so that the force becomes stabilizing and drives the mass to the origin. This fundamentally changes the physics and would be severely penalized if not corrected during the synthesizing process.

C.3 depends on C.2. The first step to solve C.3 is to copy the final result of C.2. In all 5 wrong raw solutions to C.3 (obtained conditioned on a correct solution to C.2 with the final result (\ref{eq:C2})), the copying process introduces an additional minus sign.

We suspect that such errors are a consequence of improper reinforcement learning when training Gemini 3.1 Pro Preview. In physics problems, when a force is linear in the displacement, the coefficient is usually negative (like a spring) causing harmonic oscillation. Maybe the model has seen this pattern too many times and thus tries to add a minus sign. Fortunately, when synthesizing two solutions with correct and incorrect signs, respectively, we observed that the model always produces a new solution with the correct sign.

\section{Data contamination}

Data contamination occurs when test data are intentionally or unconsciously included in a model's training data, inflating the model's apparent performance. Gemini 3.1 Pro Preview was released on February 19, 2026 with the knowledge cutoff January 2025, while the IPhO 2025 theoretical examination was held on July 21, 2025.

Some people might argue that because the exam date is after the knowledge cutoff, there is no data contamination. In our opinion, this is too optimistic. Knowledge cutoff is usually a consequence of pre-training data preparation. Training an LLM has two stages: pre-training and post-training. The former consumes an enormous amount of data (e.g., all web text) while the latter makes the model smarter by reinforcement learning on high-quality reasoning data. Since pre-training is very expensive because of its data size, it is common practice that multiple released models are post-trained on the same pre-trained model. While there is no public definition of knowledge cutoff, most people tend to believe that it refers to the last day of collecting data for the pre-training dataset. Thus, it is not unethical to include data after knowledge cutoff in the post-training dataset, and the risk of data contamination exists as long as the test dataset is released before the model.

LLMs are developing very fast, and new model versions drop every a few months. Upon model release, it is impractical to wait until the appearance of a fresh set of contamination-free datasets to evaluate it. Model release is usually accompanied with a set of benchmark results on well-known datasets that are previously public. All these evaluation results suffer from the risk of data contamination in a similar sense as our evaluation on IPhO 2025 does. Thus, avoiding or minimizing data contamination is field-wide challenge. The existence of the risk of data contamination is not a sufficient reason not to do the evaluation, but the results must be interpreted with caution. Although there are many evaluations of LLM on IPhO 2025, Ref.~\cite{QSJ+25} is the only one without data contamination and with human evaluation.

Very recently, Gemini 3 Deep Think was released and evaluated on IPhO 2025 theoretical problems. The reported accuracy is 87.7\% \cite{DT26}. It was also noted that Gemini 3.1 Pro is the ``upgraded core intelligence that makes [the Gemini 3 Deep Think] breakthroughs possible'' \cite{GT26}. This strongly suggests that Gemini 3 Deep Think is powered by Gemini 3.1 Pro. Thus, the risk of data contamination for Gemini 3 Deep Think is at the same level as that for our agent built on Gemini 3.1 Pro Preview.

Few details about the Deep Think evaluation \cite{DT26} are provided other than saying that ``Answer correctness was evaluated with reference to canonical solutions using Gemini as a judge.'' Currently, Gemini 3 Deep Think is available in the Gemini app, but to our knowledge this interface does not allow disabling web search. Gemini 3 Deep Think has not been made generally available via the Gemini API. Therefore, we cannot evaluate it on IPhO 2025 by ourselves. We have made all technical details of the evaluation of our Gemini agent fully transparent, from dataset collection and curation to human grading of the agent's solutions.

\section*{Acknowledgments}

I would like to thank Pengchuan Zhang for a discussion and Lin F. Yang for many discussions. I am especially grateful to LFY for independently confirming the error I found and described in \S\ref{ss:err}. I thank Google DeepMind for providing computational credits for using Gemini models.

\printbibliography

\end{document}